\documentclass[aps,twocolumn,showpacs]{revtex4-2}
\usepackage{graphicx}
\usepackage{amsmath}
\usepackage{amssymb}
\usepackage{color}
\usepackage{float}
\begin{document}
\title{Estimating covariant Lyapunov vectors from data}
\author{Christoph Martin}
\author{Nahal Sharafi}
\author{Sarah Hallerberg}
\affiliation{Hamburg University of Applied Sciences, Berliner Tor 21, 20099 Hamburg, Germany}
\date{\today}
\begin{abstract}
  Covariant Lyapunov vectors characterize the directions along which perturbations in dynamical systems grow.
They have also been studied as predictors of critical transitions and extreme events.
For many applications like, for example, prediction, it is necessary to estimate the vectors from data since model equations are unknown for many interesting phenomena.
We propose a novel method for estimating covariant Lyapunov vectors based on data records without knowing the underlying equations of the system.
In contrast to previous approaches, our approach can be applied to high-dimensional data-sets.
We demonstrate that this purely data-driven approach can accurately estimate covariant Lyapunpov vectors from data records generated by low and high-dimensional dynamical systems.
The highest dimension of a time-series from which covariant Lyapunov vectors were estimated in this contribution is 128.
Being able to infer covariant Lyapunov vectors from data-records could encourage numerous future applications in data-analysis and data-based predictions.
\end{abstract}
\maketitle
%
\section{Introduction}
Covariant Lyapunov vectors (CLVs) \cite{trevisan1998periodic} indicate the unstable and stable directions of dynamical systems.
As an intrinsic basis for Lyapunov exponents (LEs), they provide crucial information about the dynamical structure of a system and they form a precious tool for studying perturbation growth, coupling, and predictability in chaotic systems~\cite{pazo2010characteristic, HongLiuYang2009, Takeuchi2011, beims2016alignment, Sharafi}.
CLVs have also been employed in the context of data-assimilation for weather and climate models \cite{pazo2010spatio}.
Tangencies of CLVs are reported to be predictors for extreme events and critical transitions~\cite{Sharafi, beims2016alignment} and to be linked to climate instabilities~\cite{Lucarini2015}.

However, so far the usage of CLVs was limited to problems for which knowledge of the underlying equations of the system~\cite{wolfe, Pazo, Ginelli, Parlitz} was available, i.e., systems for which realistic models exist.
More specifically, two conditions must be addressed for the use of CLVs in data-driven scenario without knowing the underlying equations:
CLVs have to be computed without knowing the far future of the system and they have to be computed from data without the knowledge of the underlying equations.
The first issue has been addressed in~\cite{Sharafi} which proposes a method to compute approximations of CLVs without knowing the far future of the system.
A solution for   the latter problem that can be applied to multi-variate time-series of arbitrarily high-dimension (as they occur in various applications) was so far missing. 
This contribution aims to suggest a method to bridge this gap. 

One previous contribution \cite{KantzRadons} proposed to estimate CLVs through phase-space reconstruction of systems with effective dimensions of two and three. 
The method proposed in \cite{KantzRadons} is, however, not reported to be employed in applications and a generalization to high-dimensional systems was so far missing.
The reason for this is most likely the fact that phase-space reconstruction is typically limited to low-dimensional systems, since reconstruction of high dimensional systems would require enormously large data records which are typically not available in realistic application scenarios.
Recently, in a related context (estimation of optimally time-dependent modes) a combination of phase-space reconstruction and training of artificial neural networks has been suggested~\cite{Blanchard2019}.

In this contribution, we propose a novel, conceptually completely different approach for estimating CLVs from data based on the sparse identification of nonlinear dynamics \cite{Sindy}.
We demonstrate that Jacobians estimated using a modified version of SINDy can be utilized to estimate CLVs from data records of low and high-dimensional dynamical systems.

More specifically, we present the method for estimating CLVs in Sec.~\ref{sececlv} and discuss how to evaluate the quality of all estimated quantities in Sec.~\ref{specs}.
We then test our approach using trajectories of a low-dimensional chaotic system (Lorenz system)~\cite{lorenz1963deterministic} (see Sec.~\ref{secla}) , a system exhibiting critical transitions (Josephson junction)~\cite{jjp} (see Sec.~\ref{secjj}), and high-dimensional spatio-temporal chaotic systems (Lorenz 96)~\cite{lorenz1996} of dimensions $n=32$, $64$, and $128$ (see Sec.~\ref{sechd}).
We discuss the robustness of the proposed approach in the presence of added noise in Sec.~\ref{noise} and present conclusions in Sec.~\ref{secconclusions}.
\section{Estimating covariant Lyapunov vectors from data}
\label{sececlv}
Computing CLVs for a particular dynamical system
\begin{equation}
\frac{\text{d}}{\text{d}t} \mathbf{x} = \mathbf{f}(\mathbf{x},t) \label{dynsys}
\end{equation}  
requires the knowledge of the tangent operator, i.e., in numerical applications the Jacobian $\mathbb{J}_{i,j}=\dfrac{\partial f_{i}(\mathbf{x},t)}{\partial x_{j}}$, with $1 \le i,j \le n$ and $n$ denoting the dimension of the system.
Attempts to estimate the Jacobian or the leading Lyapunov exponent from data exist for a long time \cite{Sano1985,Eckmann1986, Parlitz1992, Rosenstein1993, Sauer2001, Tantet_2018}.
Here, we use a modified version of an algorithm for the sparse identification of nonlinear dynamics, (SINDy)~\cite{Sindy}.
To facilitate understanding, we first summarize the idea of SINDy.
The observed state of a system at discrete times $t=1,2,\dots, T$ is denoted by $\mathbf{x}_t$,
the matrix with all observed states by
$\mathbf{X} = [\mathbf{x}_1, \mathbf{x}_2, \dots, \mathbf{x}_T]^\top$.
The respective time derivatives of these states computed using central finite differences~\cite{fornberg1988generation} are stored in the matrix $\dot{\mathbf{X}} = [\dot{\mathbf{x}}_1, \dot{\mathbf{x}}_2, \dots, \dot{\mathbf{x}}_T]^\top$.
We then create the feature matrix (also called library) $g(\mathbf{X})$ to model the nonlinearities in the dynamics by applying several transformations to the data.
For example, $g(\mathbf{X}) = [ \mathbf{1} \text{ } \mathbf{X}  \text{ } \mathbf{X}^2 \text{ } \dots \text{ } \sin(\mathbf{X}) \text{ }  \dots ]$, with the notation referring to operations being performed elementwise.
If the underlying dynamics are unknown, the library can include features computed by many different families of functions.
The parts of the library which are effectively contributing to the dynamics are then chosen using a sparse encoding approach.
To find a sparse solution for $\dot{\mathbf{X}} = g(\mathbf{X}) \mathbf{B}$, we use a sequential threshold least-squares algorithm~\cite{Sindy}.
We refer to \cite{zhangConvergenceSINDyAlgorithm2019} for a detailed analysis of the convergence properties of this algorithm.
Other techniques for variable selection and regularization (e.g., lasso \cite{tibshirani1996regression}) could also be deployed during this step.
Note that the nonzero entries of $\mathbf{B}$ indicate the relevant terms for the dynamics of the system and thus, using $\mathbf{B}^\top g(\mathbf{x}^\top)^\top$,  we can approximate $\frac{d}{dt}\mathbf{x}$ (as in Eq.~\ref{dynsys}) for each row of the data record $\mathbf{X}$.
If the algorithm does not yield a sparse $\mathbf{B}$, the feature matrix likely does not cover the appropriate functions and thus might need to be extended. In some scenarios, other variations of the SINDy algorithm might be preferable~\cite{kahemanSINDyPIRobustAlgorithm2020}.
In applications, however, a suitable set of candidate-functions of the library has to be identified within a preselection step, by testing which candidate-functions or sets of functions provide good hindcasts of the observed data-sets.
The selected candidate functions (or their derivatives) could then form the set of functions employed in the estimation of Jacobians.

Utilizing the model derived from data, we compute numerical partial derivatives to approximate $\frac{\partial f_{i}(\mathbf{x})}{\partial x_{j}}$ which constitute the approximated Jacobian~$\hat{\mathbb{J}}$ using a central finite difference method with $4$th order accuracy~\cite{fornberg1988generation}. 
This procedure enables us to approximate Jacobians from data records without the knowledge of the underlying equations.
Section~\ref{sec:robustnesstonoise} provides an analysis of the estimation errors.
In the following, we demonstrate that these estimated Jacobians are so close to their equation-based analogues that we can use them to estimate CLVs.
This is surprising, since computing finite differences on any quantity estimated from data is typically a source of high numerical noise.
\section{Evaluating the quality of data-basted estimates (CLVs, FTLES, LEs)}
\label{specs}
We test the method proposed above by applying it to several well-known dynamical systems.
For each system under study we create data-sets of simulated trajectories which we then consider as data records.
Based on these trajectories we then estimate the Jacobians at each point in time using the procedure explained above.
The estimated Jacobians are then employed to compute several dynamical indicators: CLVs, Lyapunov exponents and finite time Lyapunov exponents (FTLEs).
We compare the CLVs computed using estimated Jacobians (in the following referred to as {\it data-based CLVs} and {\it data-based Jacobians}), to the corresponding CLVs computed using the model equations (in the following referred to as {\it equation-based CLVs}) by measuring differences in angle.
Additionally, we employ two different algorithms to compute CLVs: the algorithm of Ginelli~\cite{Ginelli}  and  the approximative near-future method (NFM)~\cite{Sharafi}.

We employ three different time steps while computing the trajectory, the Jacobian matrix and the CLVs.
In the first step we use a time step of $\delta t$ to numerically integrate Eq.~\ref{dynsys} to obtain the trajectory.
In all our models we use $\delta t = 0.0005$.
We then interprete the simulated trajectories as examples for data-records and use SINDy to estimate the Jacobians for each time-step.
In the next step, we use the estimated Jacobians to determine the evolution of 
perturbations to the trajectory using either Ginelli's algorithm or the NFM.
Therefore we use a Runge-Kutta method for iteration of the perturbation vectors and as a result the time step is increased to $2\delta t$ i.e. 0.001.
We also reorthogonalize the perturbation vectors every $\Delta t$.
In the case of the Lorenz model and the Josephson Junction $\Delta t = 0.01$, i.e. we reorthogonalize every 20 $\delta t$ steps.
As for the Lorenz 96 model $\Delta t$ is increased to 0.1. 

\begin{figure}[t!]
  \includegraphics[width=1.0\linewidth]{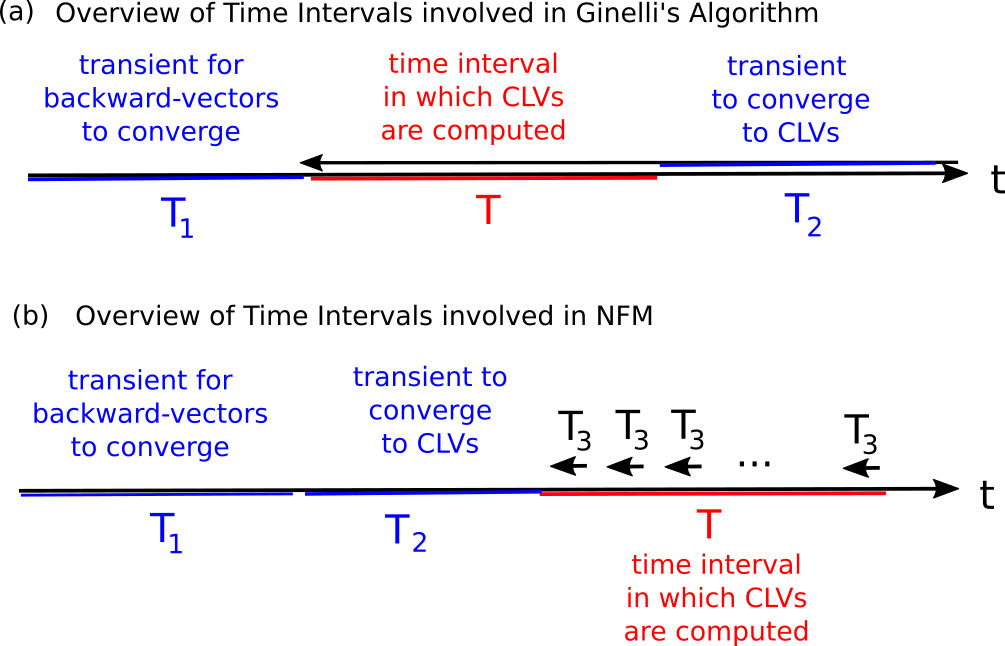}
  \vspace{-0.3cm}
 	\caption{ \label{timescetch} Scematic overviews of Time-Intervals involved in the computation of CLVs using (a) Ginelli's algorithm, (b) the near future method.}
\end{figure}
During the computation of the CLVs employing Ginelli's method we need to have a transient of length $T_1$ for forward iteration of the perturbation vectors inorder for them to converge to the backward vectors (see Fig.~\ref{timescetch} for a scematic overview of time-intervals).
We also need a transient of length $T_2$ for backward iteration of the perturbation vectors from the future. In our low-dimensional models both $T_1$  and $T_2$ are around 100.
For the Lorenz 96 model however we increased the length of both transients to 500.
For the NFM method both $T_1$ and $T_2$ are in the past.
In the case of our low-dimensional systems $T_1 = 100$ and $T_2 = 50$.
As for the case of Lorenz 96, $T_1 = 500$ and $T_2 = 100$.
The NFM then uses a short interval for backwards iterations form the future $T_3$.
This interval is 0.1 for the Lorenz attractor and Josephson Junction and $T_3  = 1$ in the Lorenz 96 model. 
We then measure the difference between data-based CLVs and equation-based CLVs generated using Ginelli's algorithm by x computing the angles $\theta_i$ between the $i$-th data-based and the $i$-th equation-based vector, with $i=1,2, \dots , n$ and $n$ denoting the dimension of the system.
Analogously, we denote the difference in angle between data-based and equation-based CLVs computed via NFM as $\phi_i$ with $i$ specified as above.
If the vectors coincide, the angle between them is zero or 180 and the absolute value of the cosine of the angle is unity.

Additionally, we compare Lyapunov exponents $l_i$ computed using data-based Jacobians and equation-based Jacobians as well as time series of the corresponding finite time Lyapunov exponents (FTLEs) $\lambda_i$, with $i=1,2, \dots,n$ as above.
\section{Results for the Lorenz attractor}
\label{secla}
\begin{figure}[t!]
  \hspace{-0.5cm}
  \includegraphics[width=1.05\linewidth]{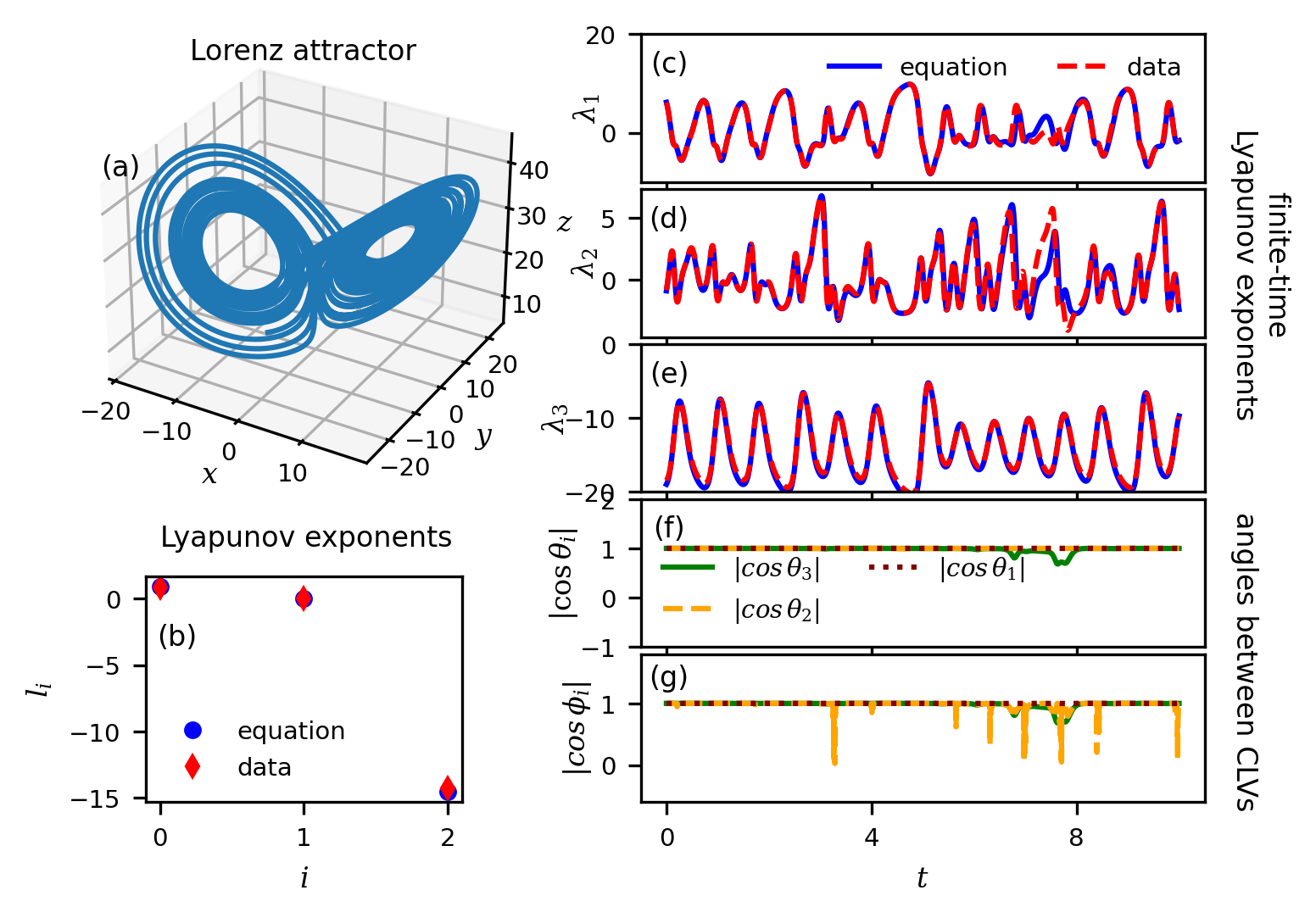}
  \vspace{-0.8cm}
 	\caption{ \label{figlosc} Approximations of CLVs, LEs, and FTLEs estimated from trajectories of the Lorenz attractor are very close to the respective quantities computed from model equations. (a)~Trajectory of the system. (b) LEs computed from model equations compared to LEs estimated from data. (c), (d), (e)~Time series of data-based estimates of FTLEs compared to  FTLEs computed from model equations. (f) Absolute value of the cosine of the angle between data-based and equation-based CLVs, both computed using Ginelli's algorithm. (g)~Similar to (f) with both sets of CLVs estimated using the NFM.}
 \end{figure}
The estimation of CLVs in low-dimensional dynamical systems is tested using 
the Lorenz attractor~\cite{lorenz1963deterministic}, the prototype model of a chaotic system given by:
\begin{equation} \label{eqlosc}
\begin{split}
\dot{x} &= \gamma(y - x)\\
 \dot{y} &= x(\rho - z) - y\\ 
 \dot{z} &= xy - \beta z\text{,}
 \end{split}
\end{equation}
with $\beta = 8/3$, $\gamma = 10$, and $\rho = 28$.
We integrate the equations of motion numerically and save the resulting Jacobians and trajectories.
The simulated trajectories (Fig.~\ref{figlosc}(a)) are in this context interpreted as {\sl data records} which we use to test our estimation procedure.
We then estimate data-based Jacobians on the basis of these data records and compute LEs, FTLEs, and CLVs using the Ginelli algorithm and the NFM.
%

%
%
As can be seen in Fig.~\ref{figlosc}(b), LEs estimated from data are very close to LEs computed using the Jacobians from differential equations of the model.
In both cases, LEs are computed using Benettin's algorithm \cite{benettin1980lyapunov} and the data-based Jacobian or the equation-based Jacobian.
The FTLEs can also be recovered from data as is presented in Figs.~\ref{figlosc}(c)--(e).   
The comparison of the angles between data-based and equation-based CLVs is presented in Fig.~\ref{figlosc}(f).
For all three CLVs, the absolute value of the cosine of the angle is most of the time very close to one, indicating that the vectors are tangent.
In Fig.~\ref{figlosc}(g), we present the corresponding results using the NFM.
Similar to Fig.~\ref{figlosc}(f), the vectors estimated from data are tangent to the ones computed using model equations ($|\cos \theta_i| \approx 1$; $i=1,2,3$).

In order to demonstrate that the time-series presented in Fig.~\ref{figlosc} refer to typical results, we also estimated histograms of angles between data-based and equation-based CLVs and histograms of FTLE estimation errors.
These histograms (see Figs.~\ref{fighistlosc} and \ref{lambdahist4}) were estimated on the basis of estimated and computed quantities for 12000 time steps and are presented in section \ref{noise} within a discussion of the robustness of the proposed method.
%
\section{Results for a fast slow system}
\label{secjj}
\begin{figure}[t!]
  \includegraphics[width=1.05\linewidth]{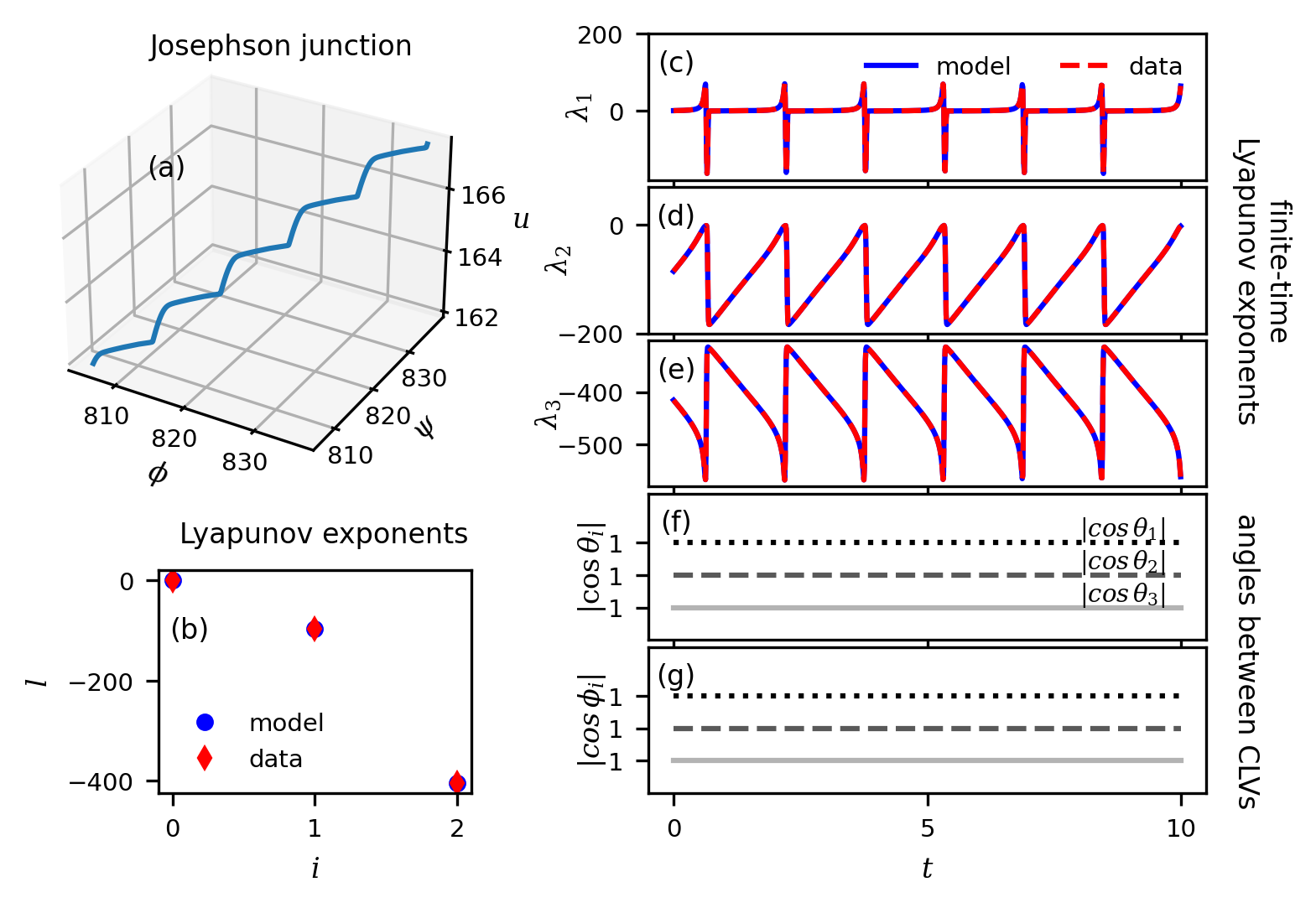}
  \vspace{-0.8cm}
 \caption{\label{figjj} Approximations of CLVs, LEs, and FTLEs estimated from trajectories of the model for Josephson junctions are very close to the respective quantities computed from model equations. (a) Trajectory of the model for Josephson junctions with transitions. (b) Data-based estimates of LEs compared to LEs computed from model equations. (c), (d), (e) Time series of data-based estimates of FTLEs compared to FTLEs computed from model equations. (f) Absolute value of the cosine of the angle between data-based and equation-based CLVs, both computed using Ginelli's algorithm. The results for the 1st and the 2nd angle are shifted by adding a constant and the labels on the y-axes are adapted in order to present all three angles in one figure. (g)~Similar to (f) with both sets of CLVs estimated using the NFM.}
\end{figure}
In order to test our approach on trajectories from a nonlinear dynamical system exhibiting critical transitions and dynamics on different time scales, we choose a model of Josephson junctions~\cite{jjb,jjp} given by:
\begin{equation} \label{eqjj}
\begin{split}
 \beta \epsilon \dot{\phi} &= \psi - (1+\beta\epsilon)\phi \\
 \epsilon \dot{\psi} &=  u - \hat{\alpha}^{-1}\phi - \mbox{sin}\, \phi\\ 
 \dot{u} &= J - \mbox{sin} \,\phi,
 \end{split}
\end{equation}
with $\hat{\alpha}^{-1} = 0.2$, $J = 1.5$, $\epsilon = 0.01$, and  $\beta = 0.2$.
This three-dimensional model has been used as a prototype model for dynamical systems with fast-slow dynamics and critical transitions~\cite{Sharafi}.

\begin{figure}[t!]
   \includegraphics[width=0.95\linewidth]{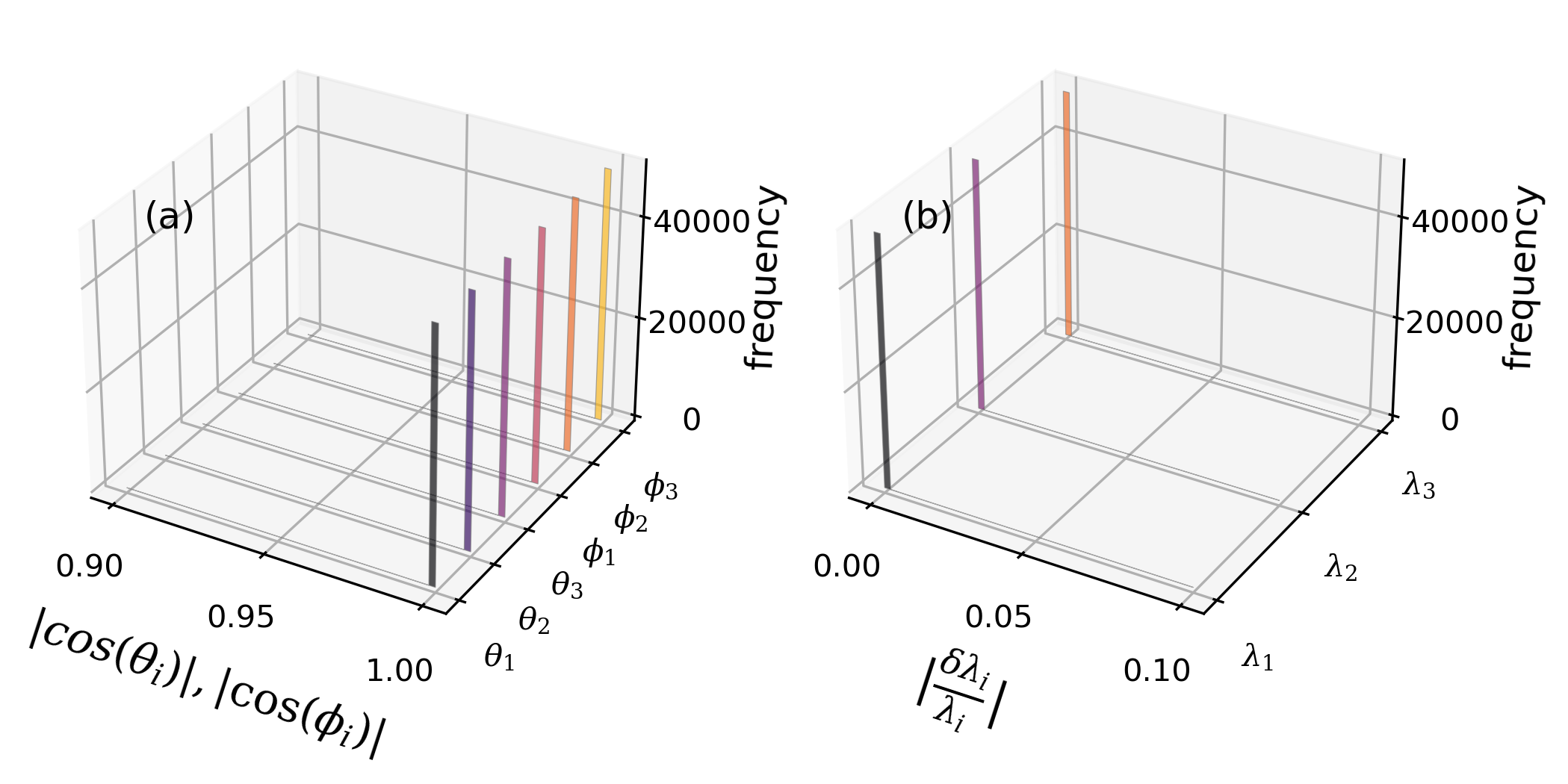}
   \vspace{-0.3cm} \caption{\label{fighistjj} The distribution of the absolute value of the cosine of the angles between CLVs from model equations and CLVs estimated from data of the Josephson Junction (b) Distribution of the error of the estimated FTLEs as compared to FTLEs from obtained from model equations. 
  }
 \end{figure}

Integrating the model equations numerically, we obtain a trajectory of a Josephson junction with transitions as is presented in Fig.~\ref{figjj}(a).
We estimate Jacobians based solely on the trajectory using SINDy and compute the LEs, FTLEs, and CLVs as explained above.
As presented in Figs.~\ref{figjj}(b)-(e), data-based estimates of LEs and FTLEs are very close to the respective equation-based quantities.
As can be seen in Fig.~\ref{figjj}(f), data-based CLVs are also tangent to equation-based CLVs,  the absolute values of the cosines of angles between data-based and equation-based CLVs are close to unity (with curves shifted to facilitate visualization and axis-labels adapted as explained in the caption). 
We observe the analogue behavior for CLVs estimated using the near future method (Fig.~\ref{figjj}(g)). 
While Figs.~\ref{figjj}(c)-(g) present only a short interval of the CLVs and FTLEs, Fig.~\ref{fighistjj} shows the distribution of the angles and relative FTLE errors estimated on the basis of estimates for 50000 time steps.
%
The distribution of the absolute value of the cosine of the angles between the data-based and equation-based CLVs is presented in Fig.~\ref{fighistjj}(a).
The sharp peaks denote that the three CLVs are always tangent using both Ginelli and NFM method.
Furthermore, Fig.~\ref{fighistjj}(b) shows the error of the data-based FTLEs as compared to the model-based FTLEs.
 Note that the absolute values of differences between estimated and equation-based FTLEs $\delta \lambda_i(t)  =  \|\hat{\lambda}_i(t) - \lambda_i(t)\|$ is divided by absolute value of the equation based FTLE, with $\hat{\lambda}_i(t)$ representing FTLEs estimated from data and $\lambda_i(t)$ denoting FTLEs computed from model equations.
 Although the histograms in Fig.~\ref{fighistjj} have entries just within one single bin, we decided to explicitly present these distributions, since they confirm that the time series presented in Fig.~\ref{figjj} represent a typical result and not a carefully-selected part of the trajectory for which the algorithm works particularly well.
%
\section{Results for a high-dimensional spatio-temporal chaotic system}
\label{sechd}
As an example for a high-dimensional multivariate chaotic time series, we simulate trajectories of Lorenz 96 models~\cite{lorenz1996} given by:
\begin{equation}
\label{eq:lorenz96}
\dot{x_i} =  -x_{i-2}x_{i-1} + x_{i-1}x_{i+1} -x_i + F ,
\end{equation}
with $F = 8$ (for this choice of the control parameter the system is chaotic), $i=1,2, \dots, n$, with cyclic indices, i.e., $x_n = x_0$ and dimensions $n=32$,  $n=64$ and $n=128$.
The resulting high-dimensional time series are then utilized to estimate Jacobians and all derived quantities.
The observed estimation errors of the Jacobians are very small.
For example, in the case of $n=128$, the mean Frobenius norm of the difference between the actual  $\mathbb{J}$ and approximated~$\hat{\mathbb{J}}$ is $3.2\text{e-}09$ with a standard deviation of $1.2\text{e-}10$.
%
%
\begin{figure}[t!]
  \hspace{-0.6cm}
  \includegraphics[width=1.05\linewidth]{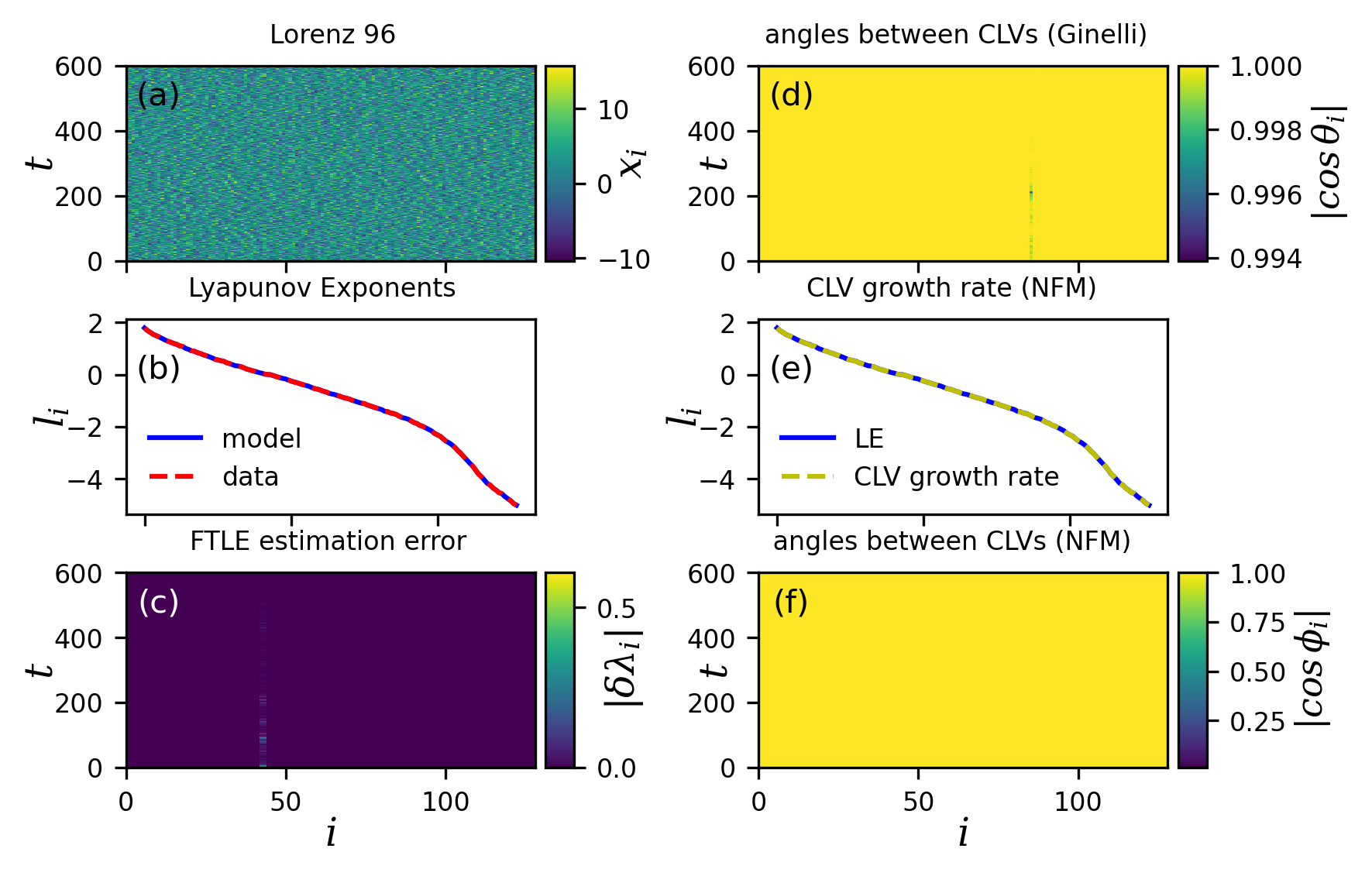}
  \vspace{-0.8cm}
 \caption{\label{figlo96} CLVs can also be estimated from time series of high-dimensional chaotic systems. Lorenz 96 system with $n = 128$ and $F = 8$. 
(a) Trajectories of the systems. (b) Data-based estimates of LEs compared to LEs computed from model equations. (c) Time-series of the differences between data-based estimates of FTLEs and equation-based FTLEs $|\delta \lambda_i(t)|$. (d) Absolute value of the cosine of the angle between data-based and equation-based CLVs. Ginelli's method has been used to compute both sets of the CLVs.  (e) LEs compared to the average growth rate of the CLVs estimated from the near future, both computed from model equations in order to test the NFM. (f) Similar to (d) with both sets of CLVs estimated using the NFM.}
\end{figure}

\begin{figure}[t!]
  \hspace{-0.6cm}
  \includegraphics[width=1.05\linewidth]{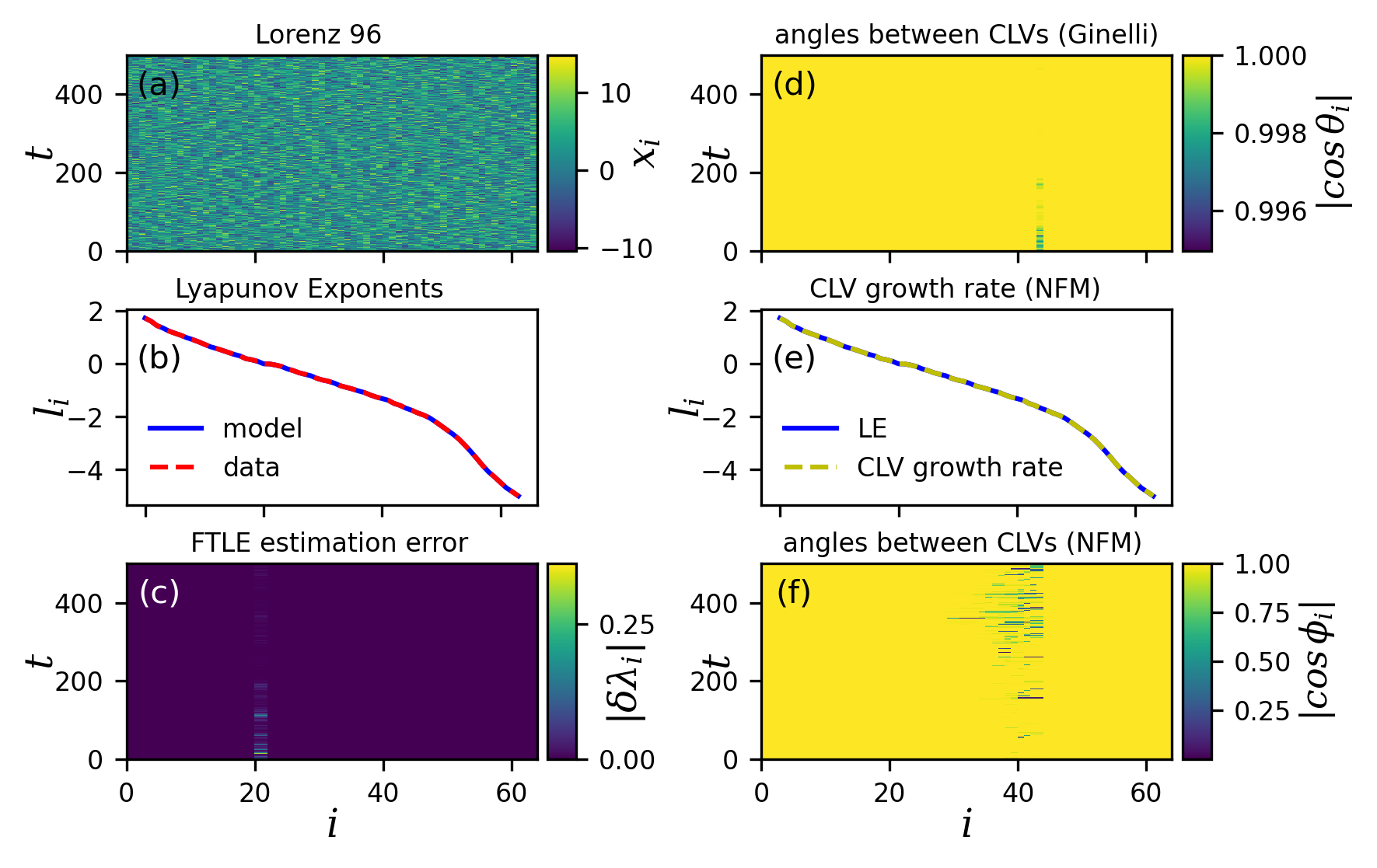}
  \vspace{-0.8cm}
  \caption{\label{fig64lo96} Data-based and equation-based quantities for a Lorenz 96 system with $n = 64$ and $F = 8$. (a) Trajectories of the systems. (b) Data-based estimates of LEs compared to LEs computed from model equations. (c) Time-series of the relative differences $|\delta \lambda_i(t)|$ between data-based estimates of FTLEs and equation-based FTLEs. (d) Absolute value of the cosine of the angle between data-based and equation-based CLVs. Ginelli's method has been used to compute both sets of the CLVs.  (e) LEs compared to the average growth rate of the CLVs estimated from the near future, both computed from model equations in order to test the NFM. (f) Similar to (d) with both sets of CLVs estimated using the NFM.}
\end{figure}

\begin{figure}[t!]
  \hspace{-0.6cm}
  \includegraphics[width=1.05\linewidth]{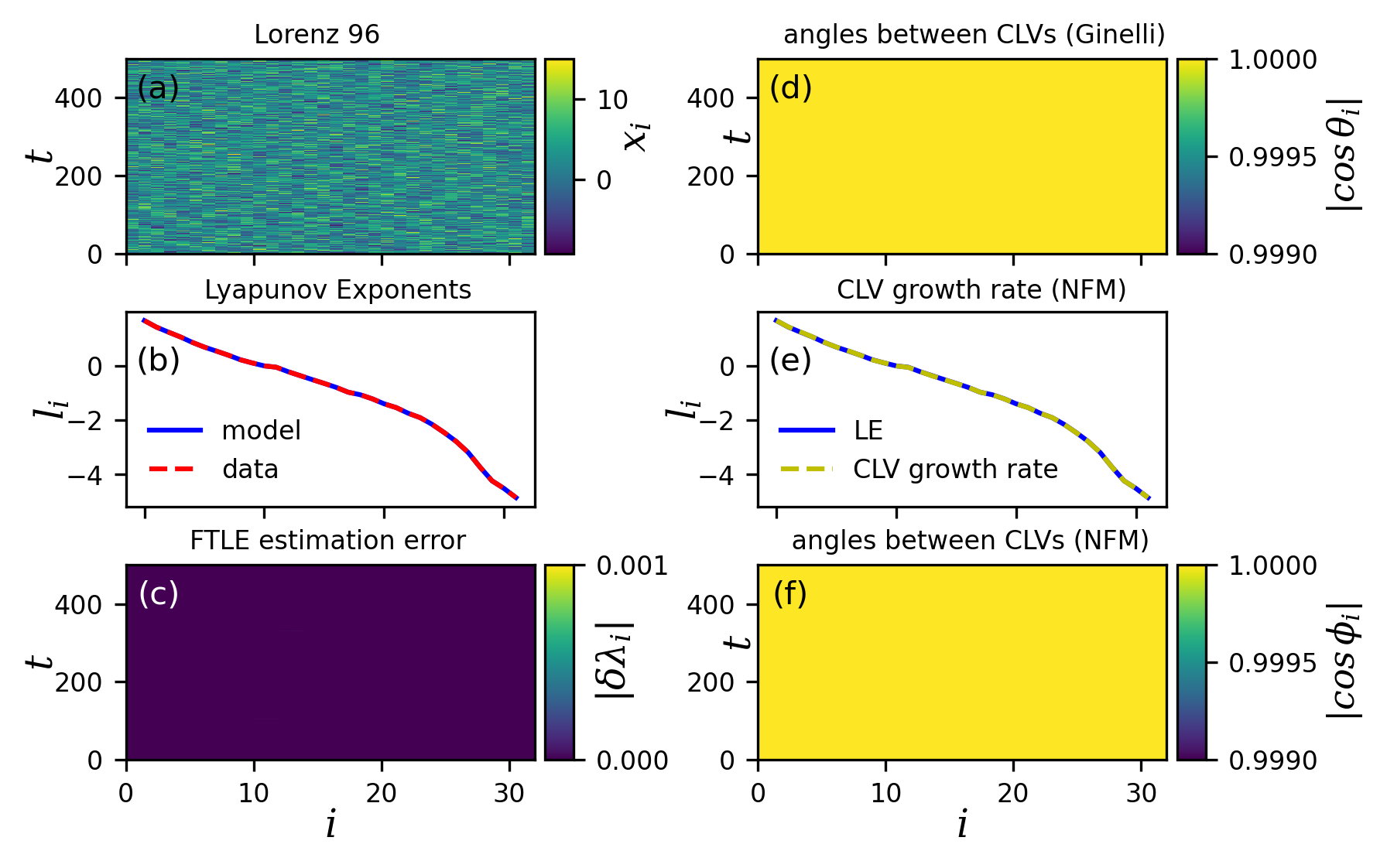}
  \vspace{-0.8cm}
  \caption{\label{fig32lo96}  Data-based and equation-based quantities for a Lorenz 96 system with $n = 32$ and $F = 8$. (a) Trajectories of the systems. (b) Data-based estimates of LEs compared to LEs computed from model equations. (c) Time-series of the relative differences $|\frac{\delta \lambda_i(t)}{\lambda_i(t)}|$ between data-based estimates of FTLEs and equation-based FTLEs. (d) Absolute value of the cosine of the angle between data-based and equation-based CLVs. Ginelli's method has been used to compute both sets of the CLVs.  (e) LEs compared to the average growth rate of the CLVs estimated from the near future, both computed from model equations in order to test the NFM. (f) Similar to (d) with both sets of CLVs estimated using the NFM.}
\end{figure}

The trajectory of a Lorenz 96 model with dimension $128$ is presented in Fig.~\ref{figlo96}(a).
Data-based estimates of the Lyapunov spectrum and the Lyapunov spectrum computed from model equations are presented in Fig.~\ref{figlo96}(b).
The estimation error of the FTLEs is presented in Figs.~\ref{figlo96}(c). 
%
%
Angles between data-based and equation-based CLVs are presented in Fig.~\ref{figlo96}(d)  using Ginelli's algorithm and Fig.~\ref{figlo96}(f) using the NFM.
The data-based CLVs are also almost tangent to the equation-based CLVs. 
To verify the effectiveness of the NFM for higher dimensions, we also compare the average growth rate of the resulting vectors to the FTLEs.
As presented in Fig.~\ref{figlo96}(e), the average growth rate of the vectors computed using the NFM coincides with the LEs and therefore these vectors can be used as a reliable basis for LEs.
Results for a Lorenz 96 models with 32 dimensions are conclusive with the 128- and 64-dimensional cases and are presented in Figs.~\ref{fig64lo96} and \ref{fig32lo96}.
%
%
%
Note that we previously observed a lower quality of the estimated vectors for intermediate values of $i$, which improved to the results presented in Figs.~\ref{figlo96}, \ref{fig64lo96} and~\ref{fig32lo96}.
More precisely, some deviations of CLVs were observed for intermediate values of $i$ and persisted in time only for a certain while before they dissolved.
Using the NFM the persistence in time was not observed, but deviations were spread among neighboring CLVs for intermediate values of $i$.
A similar but less pronounced effect is still visible in Fig.~\ref{fig64lo96}.
The quality of estimated CLVs has however, improved by increasing the length of the transients $T_1$ to the values which are specified in section \ref{specs}.
This indicates that the accuracy of estimated CLV depends on the length of the available data-records. 
We have not observed this dependence on the length of the transients for lower-dimensional systems, since convergences in lower-dimensional systems occur most likely within a shorter time interval and all time intervals we chose as transients for the lower-dimensional systems were probably larger than necessary.
%
%
\section{Testing for the robustness against noise}
\label{noise}
To verify that our approach for estimating CLVs from data still yields useful results in the presence of noise, we estimated CLVs from 
trajectories of a stochastic version of the Lorenz system.
Gaussian white noise with standard deviation $\sigma=1, 3$, and $5$ is added to all three variables in the equations of the Lorenz attractor.
When processing the trajectories of the system with noise, we apply Savitzky-Golay filtering~\cite{savitzky1964smoothing} for the reconstruction of time derivatives, before estimating Jacobians.
%
 \begin{figure}[t!]
   \includegraphics[width=1\linewidth]{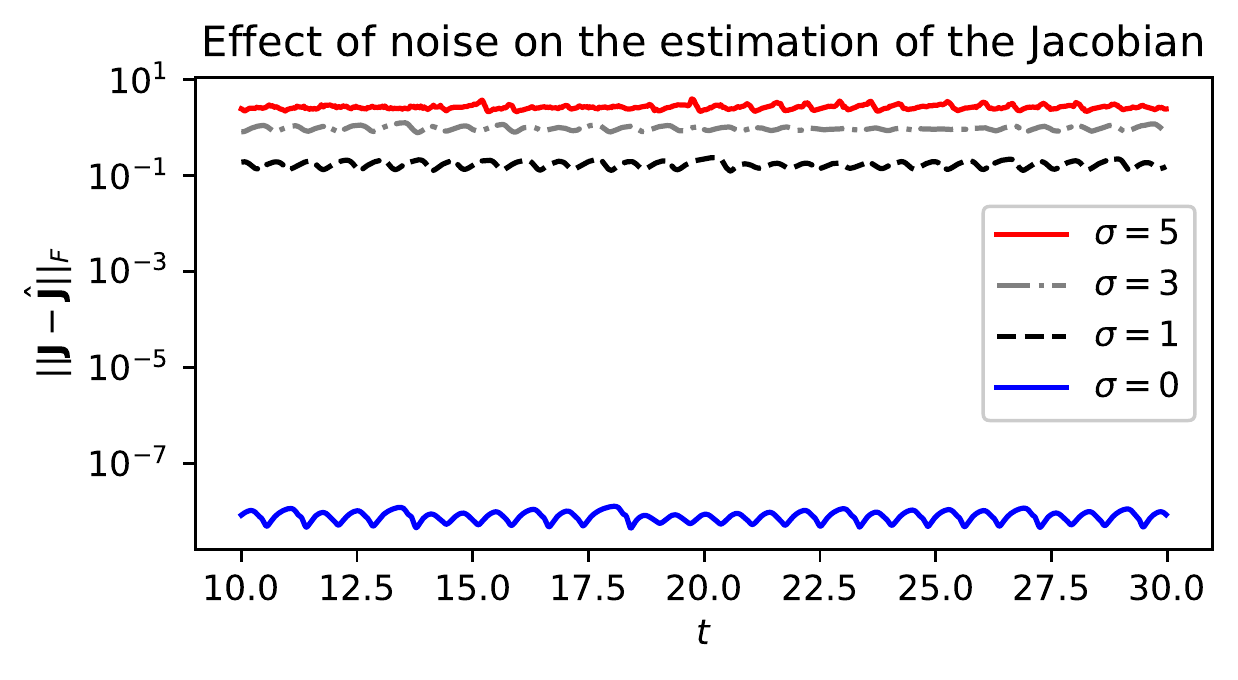}
   \vspace{-0.8cm}
  \caption{The Frobenius norm of the difference between the actual  $\mathbb{J}$ and approximated~$\hat{\mathbb{J}}$ increases with the level of noise. Note that the y-axis has a logarithmic scale. The introduction of noise has a noticeable effect on the estimation.}
  \label{fig:lorenz3dnoise}
\end{figure}
 \begin{figure}[t!]
   \hspace{-0.5cm}
   \includegraphics[width=1.05\linewidth]{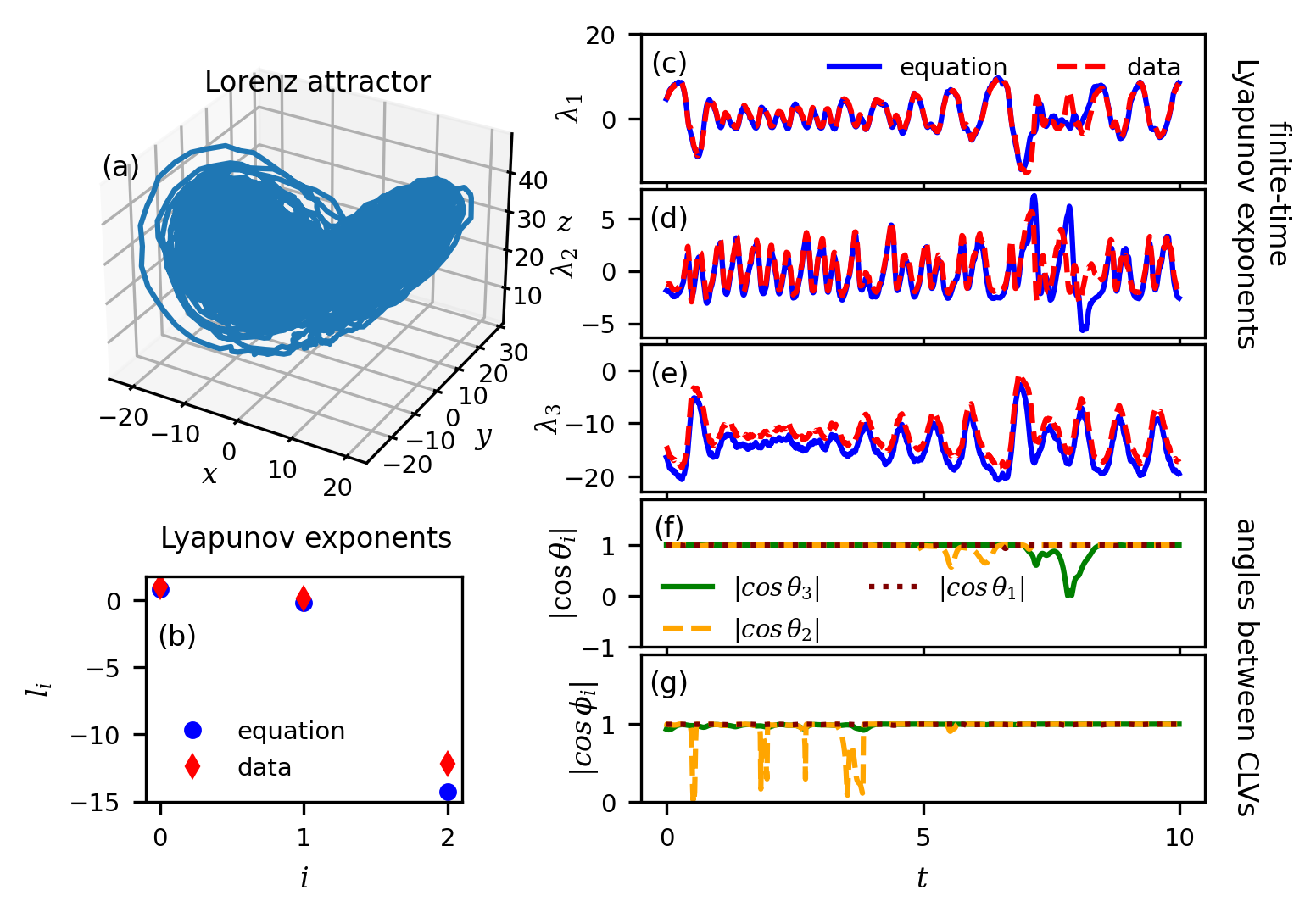}
   \vspace{-0.8cm}
  \caption{
  \label{figlosc5d} Even in the presence of noise we can obtain relatively accurate estimates of LEs, FTLEs, and CLVs. The panels above show the results for  a stochastic version of the Lorenz attractor with the standard deviation of Gaussian white noise $\sigma = 5$.
  (a) Trajectory of the system. (b) LEs computed from model equations compared to LEs estimated from data. (c), (d), (e) Time series of data-based estimates of FTLEs compared to  FTLEs computed from model equations. (f) Absolute value of the cosine of the angle between data-based and equation-based CLVs, both computed using Ginelli's algorithm. (g) Similar to (f) with both sets of CLVs estimated using the NFM.
  }
 \end{figure}
\label{sec:robustnesstonoise}
As illustrated in Fig.~\ref{fig:lorenz3dnoise}, our approach still yields useful approximations of the Jacobians, in the presence of noise.
Nevertheless, the difference between equation-based Jacobians and estimated Jacobians as measured by Frobenius norms shows an offset %
which is clearly related to the applied noise strength.
Note that the periodic fluctuations of the estimation error (best visible for $\sigma=0$) can be well explained by the changes of the FTLE shown in Fig.~\ref{figlosc}(e).
Using these approximated Jacobians based on noisy data, we compute CLVs, LEs, and FTLEs which are presented in Fig.~\ref{figlosc5d} for $\sigma = 5$.
Data-based and equation-based CLVs for the stochastic Lorenz system are still tangent most of the time, however deviations become more frequent under the influence of noise.
%

 \begin{figure}[t!]
  \includegraphics[width=0.95\linewidth]{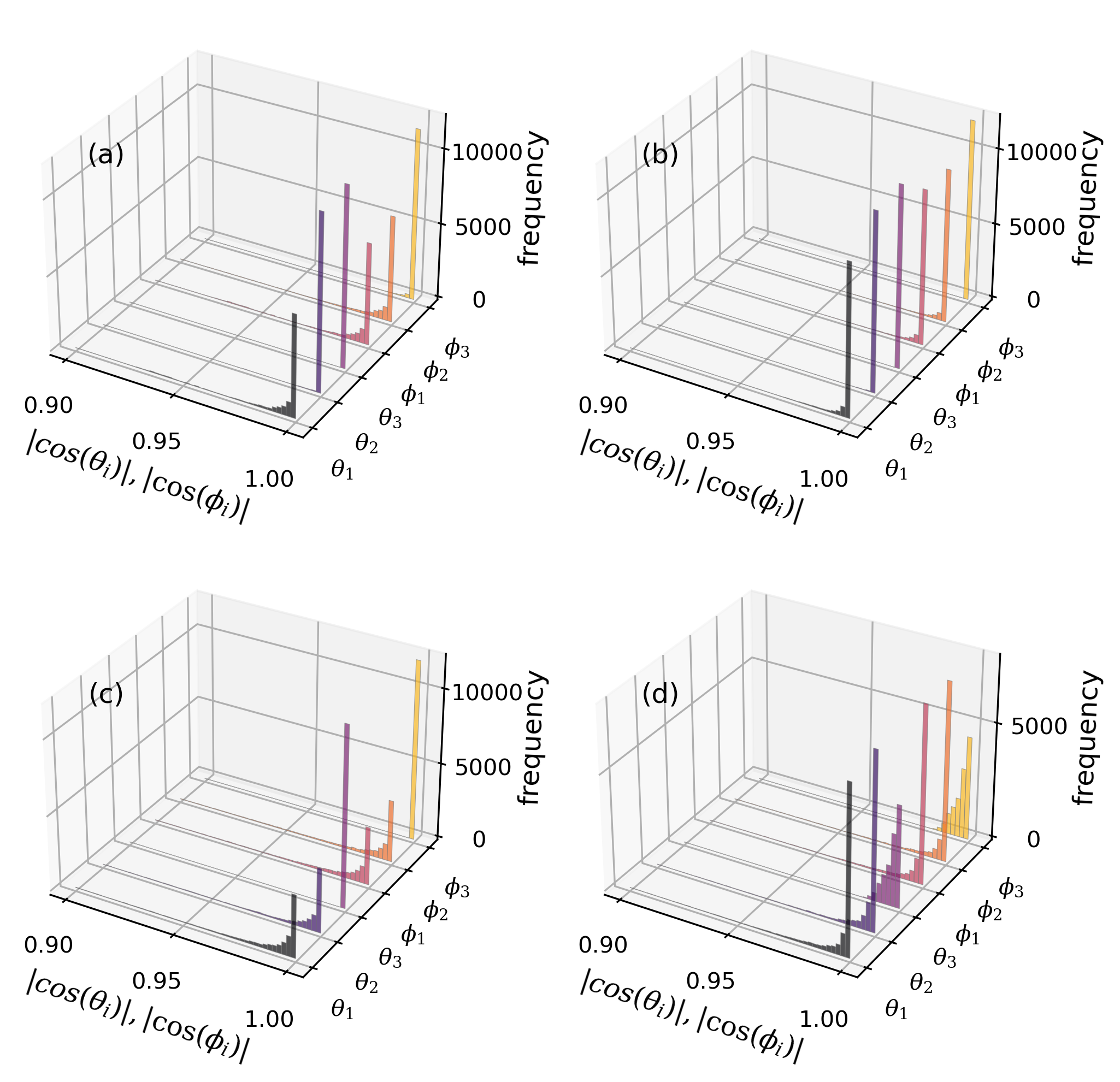}
  \caption{ 
  \label{fighistlosc} Distribution of the absolute value of the cosine of the angles between data-based CLVs and CLVs obtained using model equations in a Lorenz attractor. (a) $\sigma = 0$, (b) $\sigma = 1$, (c) $\sigma = 3$, (d) $\sigma = 5$.
  }
 \end{figure}
Additionally, we estimated histograms of the angles between data-based and equation-based CLVs, $\theta_i$ and $\phi_i$, for various noise strengths.
The resulting histograms are presented in Fig.~\ref{fighistlosc}. 
The absolute values of the cosines of angles between data-based and equation-based CLVs is unity for most time steps tested.
Deviations from this value are more frequent if the vectors are computed using the NFM method, which is expectable, since the NFM method represents an approximation to Ginelli's algorithm.
We also observe an increased spread of the mass of the distributions when the noise strength is increased.
This corresponds to the observations visible in the time series of (Fig.~\ref{figlosc5d}(f) and (g)) and the error growth of the Jacobian (see Fig.~\ref{fig:lorenz3dnoise}).

\begin{figure}[t!]
      \includegraphics[width=0.95\linewidth]{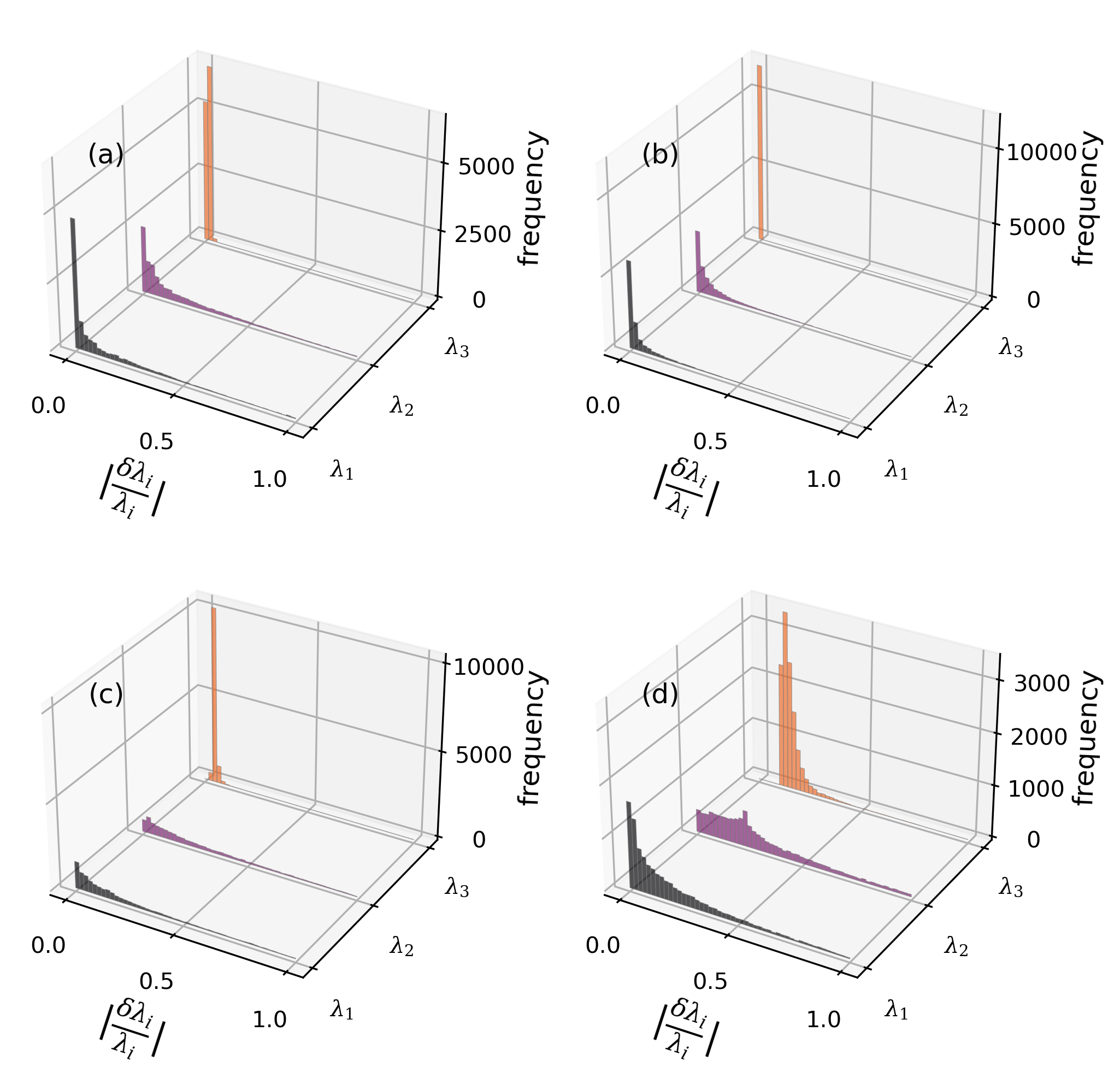}
      \caption{\label{lambdahist4} Estimation error of FTLEs for a Lorenz attractor (a) $\sigma = 0$, (b) $\sigma = 1$, (c) $\sigma = 3$, (d) $\sigma = 5$.
      }
     \end{figure}

Analogously, histograms of relative differences of data-based FTLEs and equation-based FTLEs are presented in Fig.~\ref{lambdahist4}.
The most frequent value of these differences is zero for $\sigma=0$ and $\sigma=1$. However also large differences occur (represented in the tail of the distributions), but are not very common.
If the noise is increased, the most frequent value of relative differences shifts to a very small non-zero value and larger errors become more common.
These observations correspond to the effects visible in the time series in Fig.~\ref{figlosc5d} (c)-(e) and the error growth of the Jacobian (see Fig.~\ref{fig:lorenz3dnoise}).

To summarize, we see an influence of the added noise on the quality of the estimates and this influence becomes stronger when the noise strength is increased.
Note, however that we used a relatively simple noise-filter for preprocessing (Savitzky-Golay filtering) before estimating the Jacobians and testing other, more advanced preprocessing filters could probably improve the results. 
\section{Conclusions}
\label{secconclusions}
In this contribution, we propose a novel approach for estimating CLVs of dynamical systems from data without knowing the model-equations of the system.
To obtain data-based estimates of Jacobians, we use the SINDy algorithm, a method for estimating model equations.
For testing purposes, the data records are simulations of low-and high-dimensional nonlinear systems with and without noise. 
In this proof-of-concept study, we demonstrate that the estimated Jacobians can be used to generate reliable estimates of various characteristic quantities for the analysis of local stability in dynamical systems, such as LEs, FTLEs, and CLVs.
Comparing data-based estimates with quantities computed from the equations yields results that are almost indistinguishable in the absence of noise.

Adding noise to the system leads to minor deviations between estimated LEs, FTLEs, and CLVs and their equation-based counter parts, yet the estimated quantities are still able to capture the overall dynamics of the system.
Additional preprocessing steps can easily be coupled with our approach, as we demonstrated with Savitzky-Golay filtering.
Combining our approach with more sophisticated denoising methods during the preprocessing might yield even better results.
Increasing the dimensionality of the spatio-temporal chaotic Lorenz 96 model had a minor effect on the quality of the results. 
In contrast to \cite{KantzRadons} the approach we propose is not based on phase-space reconstruction and can therefore be generalized to high-dimensional multivariate data-records.
Since the data needed to reconstruct phase-spaces scales exponentially with the dimension of the system, we can assume that the method proposed here needs considerably less data.
Note, however, that the algorithms to compute CLVs in high-dimensional dynamical systems require long-transients for the initial perturbations to convere. 
Approaches based on phase-space reconstruction can, though, address (low-dimensional) applications in which variables cannot be observed directly.
The question how missing variables would influence the quality of all estimates has not been addressed in this proof-of-concept study, but could be investigated in future contributions.

Summarizing, we propose a new method for estimating Jacobians from time series of arbitrarily high-dimensional dynamical systems and the quality of this estimates is sufficient to compute covariant Lyapunov vectors without knowing the model-equations.
Being able to estimate covariant Lyapunov vectors on the basis of data records opens up the possibility of several applications including prediction of critical transitions and studying perturbation growth on the basis of data records.

%
\section*{Acknowledgments}
The authors of this study are grateful to the BMBF for financial support within the project DADLN (01$\mid$S19079) and to the Landes\-forschungs\-f{\"o}rderung Hamburg for financial support within the project LD-SODA (LFF-FV90).

\end{document}